\documentclass[iop]{emulateapj}

\shorttitle{Substellar-mass companions to two giant stars}
\shortauthors{Gettel et al.}

\usepackage{natbib}
\begin{document}

\title{Planets Around the K-Giants BD+20 274 and HD 219415}

\author{S. Gettel\altaffilmark{1,2}, A. Wolszczan\altaffilmark{1,2}, A. Niedzielski\altaffilmark{3}, G. Nowak\altaffilmark{3}, M. Adam\'ow\altaffilmark{3}, P. Zieli\'nski\altaffilmark{3} \& G. Maciejewski\altaffilmark{3}}

\altaffiltext{1}{Department of Astronomy and Astrophysics, the Pennsylvania State University, 525 Davey Laboratory, University Park, PA 16802, sgettel@astro.psu.edu, alex@astro.psu.edu}

\altaffiltext{2}{Center for Exoplanets and Habitable Worlds, the Pennsylvania State University, 525 Davey Laboratory, University Park, PA 16802}

\altaffiltext{3}{Toru\'n Center for Astronomy, Nicolaus Copernicus University, ul. Gagarina  11, 87-100 Toru\'n, Poland, Andrzej.Niedzielski@astri.uni.torun.pl}

\slugcomment{June 4, 2012}

\begin{abstract}
We present the discovery of planet-mass companions to two giant stars by the ongoing Penn State-Toru\'n Planet Search (PTPS) conducted with the 9.2 m Hobby-Eberly Telescope. The less massive of these stars, K5-giant BD+20 274, has a 4.2 $M_{J}$ minimum mass planet orbiting the star at a 578-day period and a more distant, likely stellar-mass companion. 
The best currently available model of the planet orbiting the K0-giant HD 219415 points to a $\gtrsim$ Jupiter-mass companion in a 5.7-year, eccentric orbit around the star, making it the longest period planet yet detected by our survey. This planet has an amplitude of $\sim$18 m s$^{-1}$, comparable to the median radial velocity (RV) ``jitter'', typical of giant stars.
\end{abstract}

\section{Introduction}

Searches for planets around giant stars offer an effective way
to extend studies of planetary system formation and evolution to
stellar masses substantially larger than 1 $M_{\odot}$ \citep{sat03,sat08a,omi11b,hat06,dol09a,nie07,nie09a,qui11}. Unlike their progenitors on the main-sequence, the cool atmospheres of evolved stars produce many narrow spectral lines, making a radial velocity (RV) measurement precision of $<$ 10 m s$^{-1}$ possible.

The most conspicuous result of the RV surveys of giants is the absence of short-period planets around these stars. While hot Jupiters around main sequence dwarfs are relatively common \citep[eg.][]{wri11}, the closest planets in orbit around giants detected so far have orbital radii of $\sim$0.6 AU \citep{dol09b, nie09a}. This effect is not seen in subgiant systems, which have at least one known hot Jupiter \citep{joh10}, and is most likely related to stellar evolution \citep{vil09,nor10}. 

Typical periods for planets around giants are on the order of one year or longer, requiring multiple years of observations to characterize their orbits. The longest such period announced to date, detected by \citet{dol09a} is just over three years. In this paper, we report the detection of a planet in a nearly 6 year orbit around HD 219415. In this case, we are close to encountering the wide-orbit limit of planet detection, which results from the sensitivity of the Doppler velocity method reaching the noise level determined by the intrinsic RV jitter of giant stars.

Another consequence of the systematically lengthening time baseline of our survey is the growing frequency of planet detection in binary systems. In addition to one such case recently reported by \citet{get12}, here we describe a planet around BD+20 274, which has another, most likely low-mass stellar companion.

This paper is organized as follows. An outline of the 
observing procedure and a description of the 
basic properties of the stars are given in Section 2, followed by the 
analysis of RV measurements in Section 3. The accompanying analysis
of rotation and stellar activity indicators is given in Section 4.
Finally, our results are summarized and further discussed in
Section 5.

\section{Observations and Stellar Parameters}

Observations were made with the Hobby-Eberly Telescope \citep[HET;][]{ram98} 
equipped with the High-Resolution Spectrograph \citep{tul98}, in the 
queue-scheduled mode \citep{she07}. The spectrograph was used in the 
R = 60,000 resolution mode with a gas cell (I$_{2}$) inserted into the optical 
path, and it was fed with a 2'' fiber. Details of our survey, the observing 
procedure, and data analysis have been described in detail elsewhere 
\citep{nie07,nie08}.

The atmospheric parameters of the two stars were determined as part of an extensive study of the PTPS targets described in the forthcoming paper by \citet[][Z12]{zie12}. A brief description of the methodology employed in that work is also described in \citet{get12}. Measurements of $T_{\rm{eff}}$, log($g$), and [Fe/H] were made using the method of \citet{tak05a,tak05b}. With these values and the luminosities estimated from available data, stellar masses were derived by fitting the ensemble of parameters characterizing the
star to the evolutionary tracks of \citet{gir00}.
Stellar radii were estimated from either the log($L/L_{\odot}$) - $T_{\mbox{eff}}$ or log($g$) - $M/M_{\odot}$ relationships. Typically both methods yielded very similar values. Otherwise, the average values were computed.

The stellar rotation velocities were estimated by means of the \citet{ben84} cross-correlation method. The cross-correlation functions were computed as described by \citet{now10} with the template profiles cleaned of the blended lines. Given the estimates of stellar radii, the derived rotation velocity limits were used to approximate the stellar rotation periods. The parameters of the two stars are summarized in Table \ref{table1}.

\section{Measurements and Modeling of Radial Velocity Variations}

RVs were measured using the standard I$_{2}$ cell calibration
technique \citep{but96}. A template spectrum was constructed
from a high-resolution Fourier transform spectrometer
(FTS) I$_{2}$ spectrum and a high signal-to-noise stellar spectrum
measured without the I$_{2}$ cell. Doppler shifts were derived from
least-squares fits of template spectra to stellar spectra with the
imprinted I$_{2}$ absorption lines. Typical SNR for each epoch was $\sim$200, as measured at the peak of the blaze function at 5936 \AA. The RV for each epoch was derived
as a mean value of 391 independent measurements from
the 17 usable echelle orders, each divided into 23, 4-5 \AA\ blocks, with a typical, intrinsic uncertainty
of 6-10 m s$^{-1}$ at 1$\sigma$ level over all blocks \citep[][in preparation]{now12}. This RV precision level made it
quite sufficient to use the \citet{stu80} algorithm to refer the
measured RVs to the Solar System barycenter.

\begin{figure}               
\centering                   
\includegraphics[width=0.8\textwidth,angle=270]{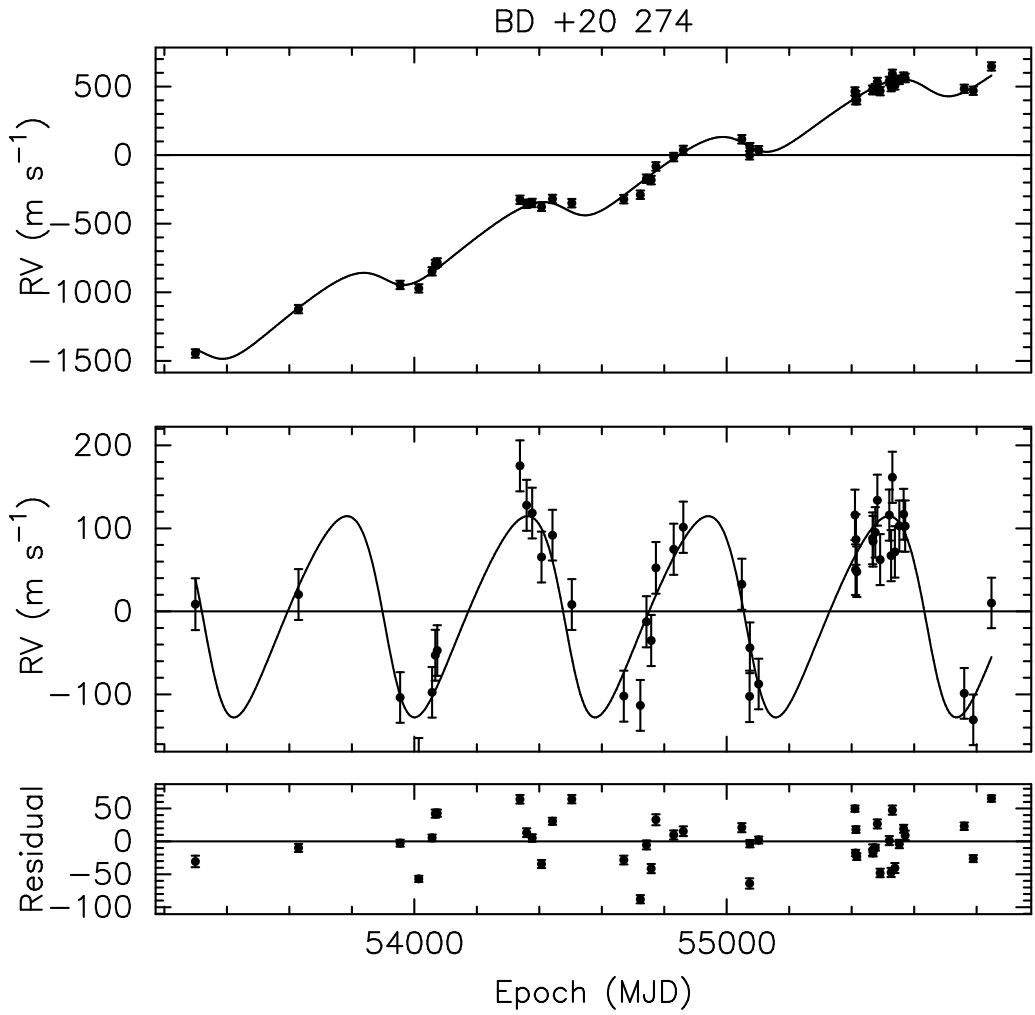}
\caption{Top: Radial velocity measurements of BD+20 274 (circles) and the best fit model consisting of the circular orbit of a long-period companion, and a 578-day planetary orbit (solid line). Middle: The RV measurements and best fit model after the subtraction of the long-term circular orbit. Bottom: The post-fit residuals for the two orbit model.  \label{fig1}}
\end{figure}

\begin{figure}               
\centering                   
\includegraphics[width=0.8\textwidth,angle=270]{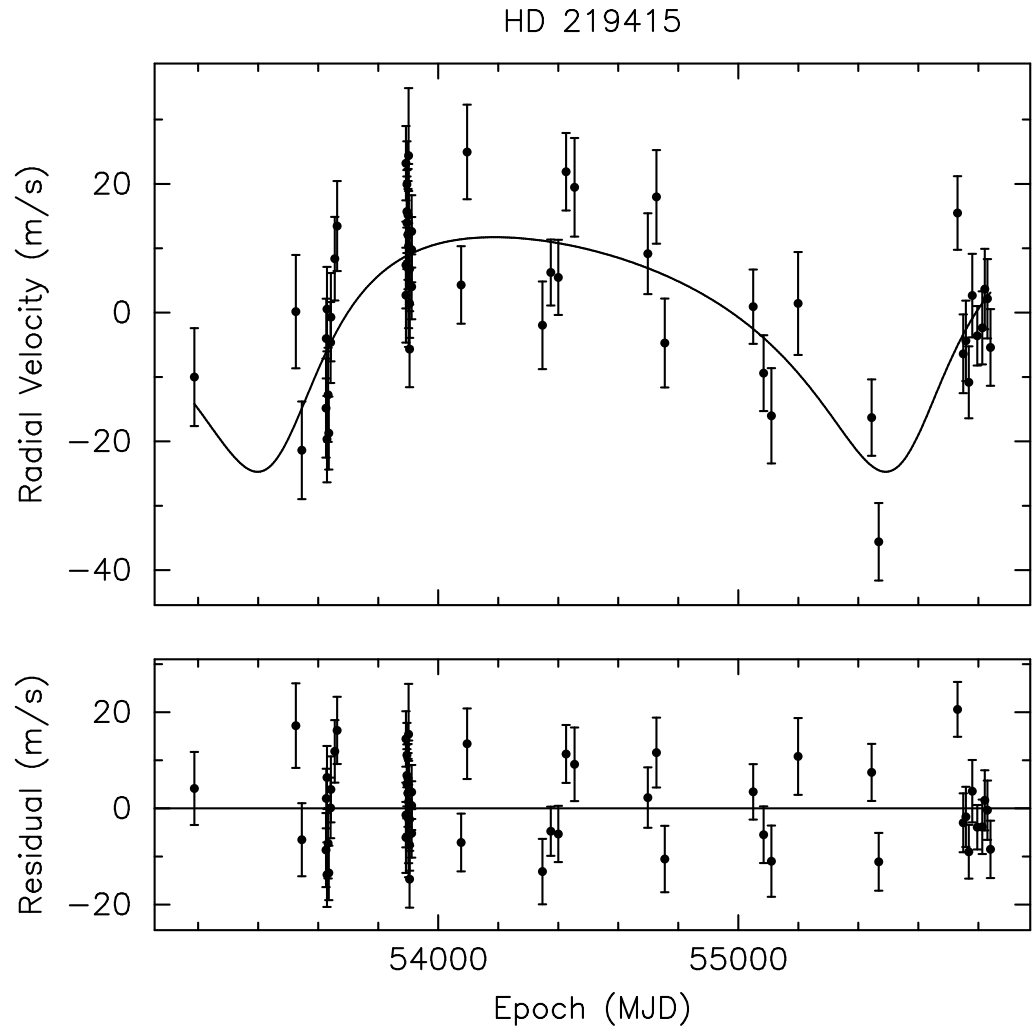}
\caption{Top: Radial velocity measurements of HD 219415 (circles) and the best-fit of a single planet Keplerian model (solid line). Bottom: The post-fit residuals for the single planet model.  \label{fig2}}
\end{figure}

The RV measurements of each star were modeled in terms of the standard,  
six-parameter Keplerian orbits, as shown in Figures \ref{fig1} and \ref{fig2}. Least-squares fits to the data were performed using the Levenberg-Marquardt algorithm \citep{pre92}. Errors in the best-fit orbital parameters were estimated from the parameter covariance matrix. These parameters are listed in Table \ref{table2} for each of the two stars. In the case of BD+20 274, the estimated, $\sim$7 m s$^{-1}$ errors in RV measurements were evidently too small to account for the actually measured post-fit rms residuals (Table \ref{table2}), whereas for HD 219415 the two quantities had comparable values. 

Giant stars are subject to atmospheric fluctuations and radial and non-radial pulsations that may manifest themselves as an excess RV variability. In particular, the p-mode oscillations
can account for a significant fraction of the observed post-fit RV noise \citep{hek06}. We estimate the amplitude of such variations using the scaling relations of \citet{kje95,kje11}. 

The statistical significance of each detection was assessed by calculating false alarm probabilities (FAP) using the RV scrambling method \citep[][and references therein]{wri07}. The FAP for HD 219415 was calculated for the null hypothesis that the planetary signal can be adequately accounted for by noise. The FAP for BD+20 274 was calculated for the null hypothesis that it can be modeled as a fraction of a long-period, circular orbit and noise. 

\subsection{BD+20 274}

RVs of BD+20 274 (AG +20 153) are listed in Table \ref{table3}. They were measured at 43 epochs over a
period of 2550 days from October 2004 to October 2011. The SNR values ranged from 170 to 312. The exposure time was selected according to actual weather conditions and ranged between 527 and 824 s.
The estimated median RV uncertainty for this star was 6 m s$^{-1}$. 

Radial velocity variations of BD+20 274 over a 7-year period are
shown in Figure \ref{fig1}. They are characterized by a long-term upward trend, with the superimposed periodic variations. The best-fit Keplerian orbit model for the periodic component, with the trend modeled as a circular orbit, is given in Table \ref{table2}. For the assumed stellar mass of 0.8 $M_{\odot}$, the planetary companion has a minimum mass of $m_{2}$ sin $i$ = 4.2 $M_{J}$, in a 578-day, mildly eccentric orbit with a semi-major axis of 1.3 AU. The FAP for this signal is $<$ 0.01\%.

Characterization of the orbit of the more distant companion is difficult, as its period is much longer than the time baseline of our measurements. As there is little curvature to the observed trend, the most straightforward approach is to use the measured linear RV drift rate, $\dot V\sim$0.81 m s$^{-1}$day$^{-1}$, to approximate the companion mass (in $M_{J}$) as a function of orbital radius from \citep{bow10}: 
\begin{equation}
m_2\approx 1.95\frac{\dot V}{(1 m~s^{-1}~day^{-1})}\bigg(\frac{a^2}{1 AU}\bigg) 
\end{equation}

This shows that, for $a \geq 3$ AU  the outer companion becomes a brown dwarf and at $a \geq 7$ AU the outer companion has a mass of 80 $M_{J}$, becoming another star.  These limits can be further restricted with the $\sim$0.4 M$_{\odot}$ minimum companion mass derived from the above provisional fit of a partial circular orbit to the RV data for BD+20 274.


The post-fit residuals for this model exhibit a large, 35 m s$^{-1}$ rms noise. Assuming that this RV jitter is due to solar-like oscillations, the Kjeldsen-Bedding relation predicts an amplitude of $\sim$27 m s$^{-1}$ for this star. To account for these variations, we have quadratically added 30 m s$^{-1}$ to the RV measurement uncertainties before performing the least-squares fit of the orbit.

\subsection{HD 219415}

RVs of HD 291415 (BD+55 2926) were measured at 57 epochs over a
period of 2650 days from July 2004 to October 2011 (Table \ref{table5}). The SNR for these measurements ranged from 150 to 290. The exposure time ranged between 268 and 600 s.
The estimated median RV uncertainty for this star was 
6.4 m s$^{-1}$.

The radial velocity variations of HD 219415 over a 7-year period are
shown in Figure \ref{fig2}, together with the best-fit model of a Keplerian orbit \ref{table3}. At 5.7 years, this orbit has the longest period of any planet detected by our survey. For the assumed stellar mass of 1.0 $M_{\odot}$, the planetary companion has a minimum mass of $m_{2}$ sin $i$ = 1.0 $M_{J}$, in a 3.2 AU, eccentric orbit. The FAP for this signal is $<$ 0.01\%.

The Kjeldsen-Bedding amplitude estimate for this star is 1 m s$^{-1}$. 
Because the post-fit residual of 9 m s$^{-1}$ rms RV jitter is comparable to the formal measurement error, there was no need to increase it to account for the intrinsic RV variations.

\section{Stellar Photometry and Line Bisector Analysis}

Photometry and line bisector analysis, which are efficient stellar activity indicators, have been routinely used to verify the authenticity of planet detections by means of the RV method \citep[e.g.][]{que01}. As simultaneous photometric data are typically not available for our targets, it is useful to correlate the time variability of line bisectors and RVs, in order to obtain additional information on a possible contribution of stellar activity to the observed RV behavior. For example, even for slow rotators like the two stars discussed here, a stellar spot only a few percent in size could introduce potentially detectable line profile variations \citep{hat02}. On the other hand, one must bear in mind that, in the case of these stars, the instrumental profile width of 5 km s$^{-1}$ is comparable to the rotational line broadening (Table 1), which restricts the diagnostic value of line bisector analysis.

\begin{figure}
\centering
\includegraphics[width=0.8\textwidth,angle=270]{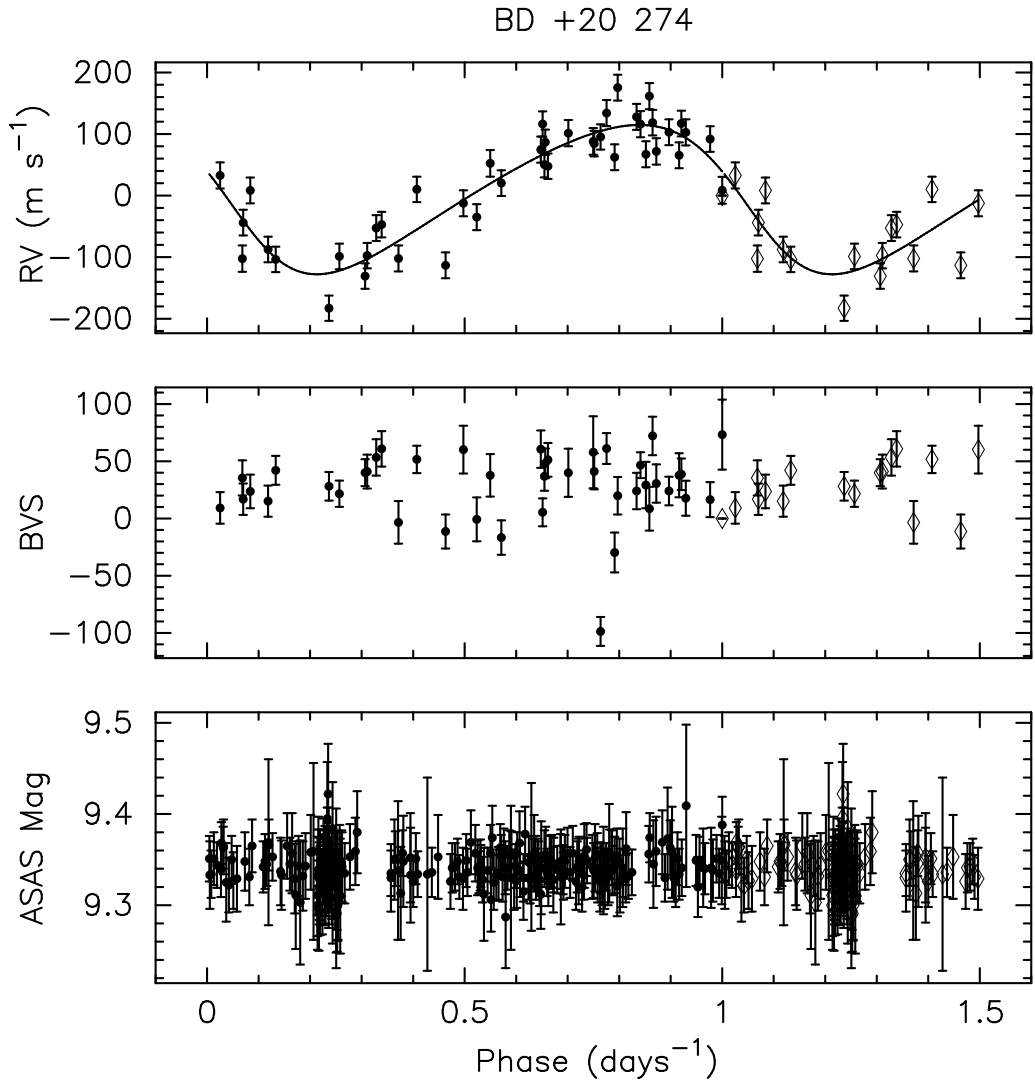}
\caption{BD+20 274 data folded at the best-fit orbital period. From top to bottom: (a) Radial velocity measurements after the removal of the outer orbit, along with the best fit inner planet model (b) Bisector Velocity Span (m/s) (c) ASAS photometry.\label{fig3}}
\end{figure}
\begin{figure} 
\centering
\includegraphics[width=0.8\textwidth,angle=270]{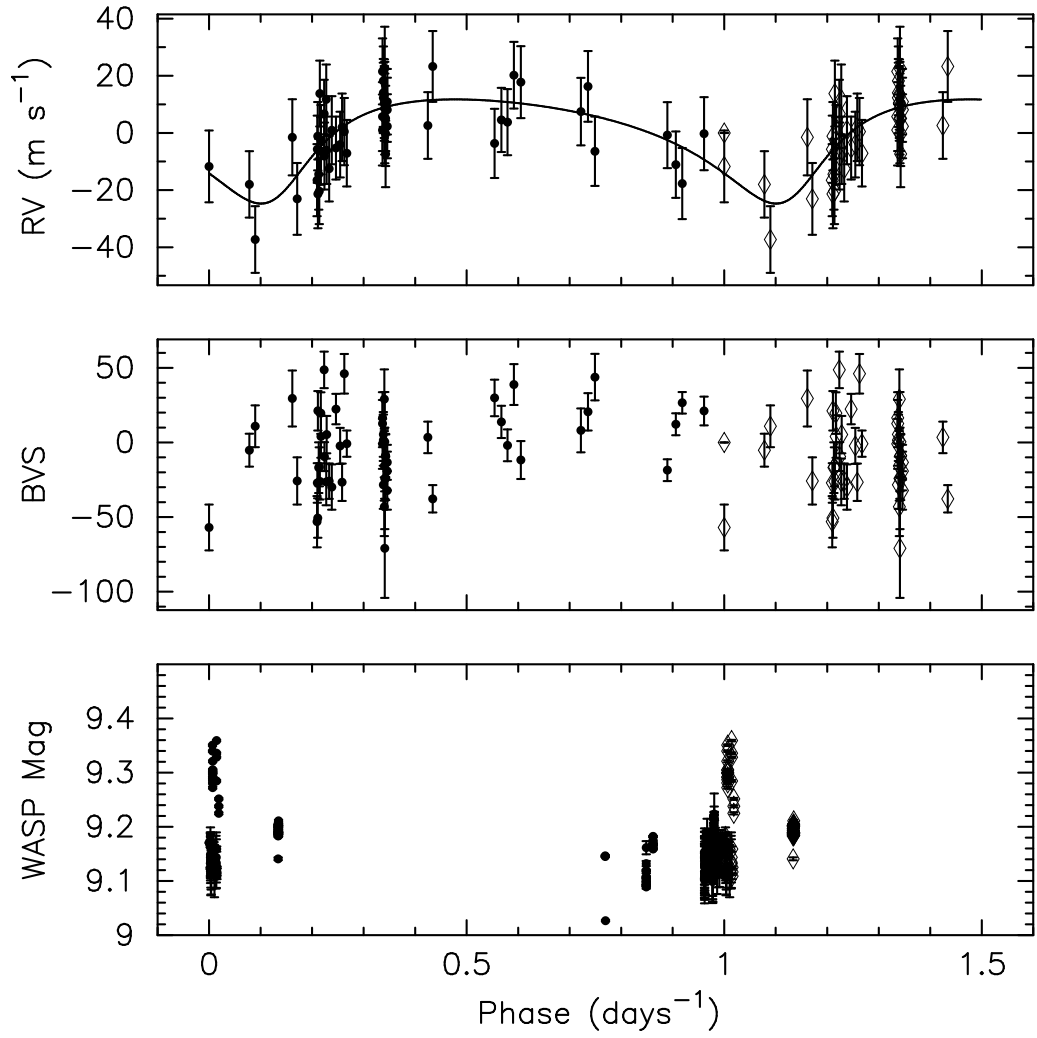}
\caption{HD 219415 data folded at the best-fit orbital period. From top to bottom: (a) Radial velocity measurements with the best fit model (b) Bisector Velocity Span (m/s) (c) WASP photometry.\label{fig4}}
\end{figure}

In order to investigate a possible contribution of stellar jitter to the observed RV periodicities, we have examined the existing photometry data in search for any periodic light variations, and performed a thorough analysis of time variations in line bisector velocity span (BVS) for the two stars. In addition, assuming that the scatter seen in the photometry data is solely due to a rotating spot, we have used it to estimate the corresponding amplitude of RV and BVS variations as in \cite{hat02}. The BVSs were measured using the cross-correlation method proposed by \citet{mar05} and applied to our data as described in \citet{now10}. For each star and each spectrum used to measure RVs at all the observing epochs, cross-correlation functions (CCFs) were  computed from $\sim$1000 line  profiles with the I$_{2}$ lines removed from  the spectra. The time series for these parameters and the photometric data folded at the observed RV periods for the two stars are shown in Figures \ref{fig3} and \ref{fig4}. 


\subsection{BD+20 274}

The photometry available for this star
consists of 243 ASAS  measurements \citep{poj02} spanning the period between
MJD 52625 and 55166. These data give a mean magnitude of the star of $V = 9.34 \pm 0.02$. There  are also 44  epochs of usable photometric observations  of this
star available from the Northern Sky Variability Survey \citep[NSVS;][]{woz04} with the mean $V=9.11\pm0.02$. Neither of these time series exhibit any periodic brightness variations. 

The calculated mean value of the BVS  for this star is  $28.3 \pm
30.7$ m s$^{-1}$. No correlation between the RV variations and those of the BVS was found with a correlation coefficient of r = -0.14.   
The expected RV and BVS amplitudes due to a possible presence of a spot, estimated from the photometric scatter, are 30 m s$^{-1}$ and 6 m s$^{-1}$, respectively, which is much less than the observed variability.


 

 
\subsection{HD 219415}

Although the photometric data from two sources exist for HD 219415, neither offers a sufficient phase coverage of this very long orbit. We have used 218  WASP measurements \citep{pol06} made between MJD 54388 and 54312, to obtain a mean magnitude of the star of $V = 9.17 \pm 0.05$. The additional 59 measurements from NSVS averaged at  $V=8.93 \pm 0.03$. Again, neither of these data sets exhibit periodic brightness variations.

The mean value of the BVS for this star is $-4.7 \pm 26.8$ m s$^{-1}$ and a correlation between the RV and the BVS is not significant, with a correlation coefficient of r = -0.01.  
The photometry derived RV and BVS amplitudes due to a presence of a spot are 21-34 m s$^{-1}$ and 23-36 m s$^{-1}$, respectively and are comparable to the observed RV variability. However, if the radial velocity variations of HD 219415 were caused by a starspot, the period of the signal would be approximately equal to the stellar rotation period.  While our period estimate of $\sim$140$\pm$1200 days is highly uncertain, it is clearly inconsistent with the observed RV variations of the star.

\section{Discussion}

In this paper, we report detections of periodic RV variations in two K-giant stars from the list of targets monitored by the PTPS program. The accompanying analyses of the photometric data, as well as of the time variability of line bisectors indicate that the most likely origin of the observed periodicities is the Keplerian motion of planet-mass companions. 

The selection criteria of our survey have been designed to reject targets that show a RV ramp of $>1$ km s$^{-1}$ in preliminary measurements over the period of 2-3 months. However, stars with a ramp of $<1$  km s$^{-1}$ over that time are continued to be observed, if they exhibit $>20$ m s$^{-1}$ RV scatter around this trend. This has the net effect of rejecting close binaries, but allowing the detection of planets in wide binary systems. Indeed, several of the planetary systems detected by PTPS, including BD+20 274, show long-term radial velocity trends with linear drift velocities of order 1 m s$^{-1}$ day$^{-1}$, suggesting that these stars have binary companions. While the dynamical constraints from these partial orbits are not highly restrictive, further information about the secondaries can be obtained by examining the stellar spectra. 

The spectra of BD+20 274 do not show signs of lines from a second star, even in the stellar template which has S/N = 355. If we assume that the companion must have S/N $\sim$10 to be detectable, this suggests that the luminosity ratio between the two objects is of order $\sim$1300, or about 8 magnitudes. As the absolute magnitude of a typical giant star is $M_{bol} = 0.08$, the maximum mass of a MS companion is 0.5 M$_{\odot}$. 
By examining the cross-correlation function used in the line bisector measurements in the manner of \citet{que95}, this estimate can be further restricted. Using the many lines in the CCF template increases the signal, giving a detectable line intensity ratio for the two stars of $\sim 3200$. This corresponds to a secondary that is $\sim$9 magnitudes fainter, or about 0.3 M$_{\odot}$ for a main sequence star. Both these limits are consistent with those discussed in Section 3. 

The range of orbital radii of planets discoverable around giants is restricted by stellar evolution and the $\sqrt{a}$ minimum mass scaling of the Doppler velocity method. Indeed, no planet around a K-giant with an orbital radius smaller than 0.6 AU has been detected so far \citep{dol09b, nie09a}, in general agreement with the theoretical estimates based on the influence of tidal effects and stellar mass-loss on orbital evolution \citep{vil09,nor10,kun11}. At the other end of the range of orbital sizes, the HD 219415 planet reported in this paper illustrates the detectability limits of wide orbit, long-period planets imposed by the enhanced intrinsic RV jitter in giants. This effect has been studied by \citet{hek06}, who have demonstrated that the rms RV noise of K-giants has a median value of $\sim$20 m s$^{-1}$ and it tends to increase toward later spectral types. 

\begin{figure}
\centering
\includegraphics[width=0.8\textwidth]{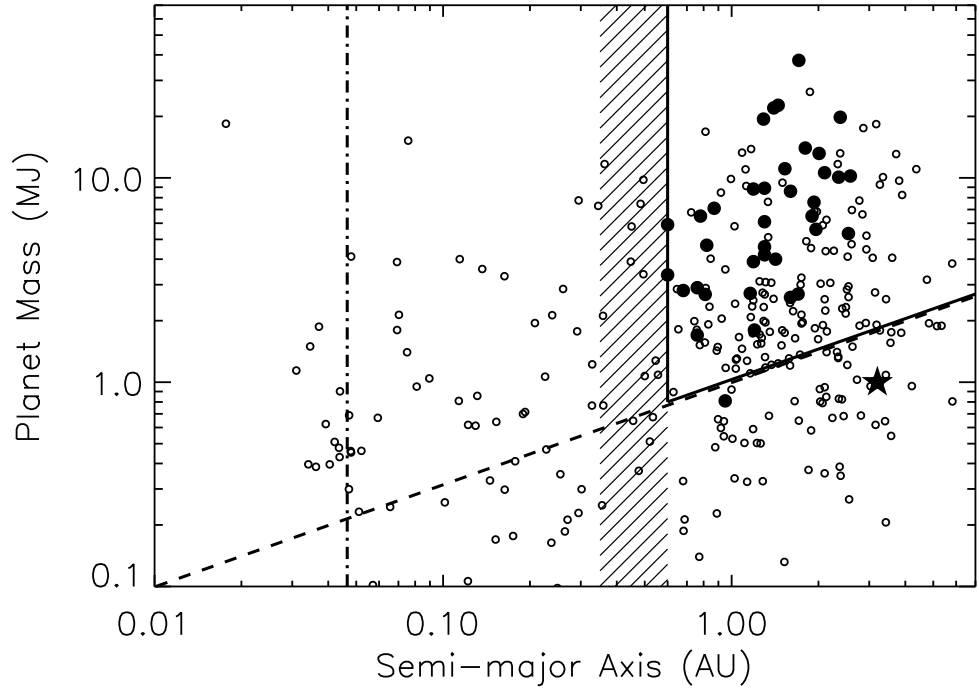}
\caption{Discovery space for RV planets around dwarfs (open circles) and giant stars (filled circles). A detection sensitivity for typical jitter amplitudes in early giants ($\sim$20 m s$^{-1}$)with an assumed mass of 2 M$_{\odot}$ is indicated by the dashed line. The shaded region marks the range of minimum orbital radii predicted by theory. The dash-dotted line at 10 R$_{\odot}$ marks the typical radius of an early giant. The solid lines mark the discovery space for planets around giants. The star marks HD 219415.\label{fig5}}
\end{figure}

Both these limits are outlined in Figure \ref{fig5}. Evidently, planets down to the Saturn-mass should be easily detectable around early giants over the 0.1-0.5 AU range of orbital radii, but their existence is apparently impaired by the dynamical effects of stellar evolution. For wide orbits, a 1 M$_{J}$ planet around a 2 M$_{\odot}$ star would have a RV signal of $<$20 m s$^{-1}$ for $a \geq$ 1 AU, and could be buried in the intrinsic RV noise of a late-type giant. This detection threshold will become more important as the time baseline of the ongoing giant surveys continues to expand, and it will eventually place a practical upper limit on long-period planet detection around the evolved stars.

\bigskip
We thank the HET resident astronomers and telescope operators for support. SG and AW were supported by the NASA grant NNX09AB36G.  AN, MA, PZ and GN were supported in part by the Polish Ministry of Science 
and Higher Education grant N N203 510938 and  N N203 386237. GM acknowledges the financial support from the Polish Ministry 
of Science and Higher Education through the Juventus Plus grant IP2010 
023070.

The HET is a 
joint project of the University of Texas at Austin, the Pennsylvania State University, Stanford University, Ludwig-Maximilians-Universit\"at M\"unchen, and Georg-August-Universit\"at G\"ottingen. 
The HET is named in honor of its principal benefactors, William 
P. Hobby and Robert E. Eberly. The Center for Exoplanets and 
Habitable Worlds is supported by the Pennsylvania State University, the Eberly College of Science, and the Pennsylvania 
Space Grant Consortium. 

\clearpage
\bibliographystyle{apj}
\bibliography{disc2}

\clearpage
\newpage

\begin{deluxetable}{lcc}
\tablewidth{200pt}
\tablecolumns{3}
\tablecaption{Stellar Parameters\label{table1}}
\tablehead{\colhead{Parameter} & \colhead{BD+20 274} & \colhead{HD 219415} }
\startdata
V & 9.36 & 8.94\\
B-V & 1.358 $\pm$ 0.067 & 1.002 $\pm$ 0.027\\
Spectral Type & K5 III & K0 III\\
$T_{\rm{eff}}$[K]& 4296 $\pm$ 10 & 4820 $\pm$20\\
log($g$) & 1.99 $\pm$ 0.05 & 3.51 $\pm$ 0.06\\
$$[Fe/H] & -0.46 $\pm$ 0.07 & -0.04 $\pm$ 0.09\\
log($L_{\star}/L_{\odot}$) & 1.96 $\pm$ 0.18 & 0.62 $\pm$ 0.15\\
$M_{\star}/M_{\odot}$ & 0.8 $\pm$ 0.2 & 1.0 $\pm$  0.1\\
$R_{\star}/R_{\odot}$ & 17.3 $\pm$ 0.9 & 2.9 $\pm$ 0.4\\
$V_{rot}$[Km/s] & 2.0 $\pm$  1.5 & 1.1 $\pm$ 1.5 \\
\enddata
\end{deluxetable}

\begin{deluxetable}{lcc}
\tablecolumns{3}
\tablewidth{200pt}
\tablecaption{Orbital Parameters\label{table2}}
\tablehead{\colhead{Parameter} &  \colhead{BD+20 274 b}  & \colhead{HD 219415 b} }
\startdata
P [days] & 578.2$\pm$5.4 & 2093.3$\pm$32.7\\
$T_{0}$ [MJD] & 53920.5$\pm$33.2 & 53460.9$\pm$57.7\\
K [m/s] & 121.4$\pm$6.4 & 18.2$\pm$2.2 \\
e & 0.21$\pm$0.06 &0.40$\pm$0.09\\
$\omega$ [deg] & 108.5$\pm$38.1 &207.0$\pm$12.3 \\
$m_{2}sin(i)$ [$M_{J}$]&4.2 &1.0 \\
$a$ [AU] & 1.3 &3.2\\
$\chi^{2}$ &1.7 &2.0 \\
Post-fit rms [m/s] & 35.8 & 8.8 \\
\enddata
\end{deluxetable}

\begin{deluxetable}{lcc} 
\tablecolumns{3}
\tablewidth{200pt}
\tablecaption{Relative RVs of BD+20 274 \label{table3}}
\tablehead{\colhead{Epoch [MJD]} & \colhead{RV [m s$^{-1}$]} & \colhead{$\sigma_{RV}$ [m s$^{-1}$]}}
\startdata
   53299.17090   &    -1446.2    &      8.7     \\  
   53629.46776   &    -1122.8    &      6.2     \\  
   53954.38543   &     -946.8    &      5.3     \\  
   54014.22433   &     -971.8    &      4.7     \\  
   54057.10443   &     -847.7    &      5.1     \\  
   54067.27214   &     -793.8    &      6.2     \\  
   54073.24994   &     -782.8    &      5.6     \\  
   54338.32091   &     -325.7    &      6.5     \\  
   54359.47756   &     -354.9    &      6.6     \\  
   54377.41906   &     -348.7    &      5.7     \\  
   54407.14137   &     -376.1    &      6.1     \\  
   54442.05559   &     -319.7    &      5.4     \\  
   54504.07923   &     -350.3    &      6.3     \\  
   54670.41663   &     -321.2    &      6.7     \\  
   54723.28207   &     -288.9    &      6.2     \\  
   54743.23171   &     -171.8    &      6.9     \\  
   54758.18574   &     -182.2    &      7.0     \\  
   54773.34312   &      -82.4    &      8.3     \\  
   54830.18984   &      -14.2    &      7.0     \\  
   54861.11542   &       37.2    &      7.3     \\  
   55048.39243   &      115.1    &      6.8     \\  
   55073.31980   &       -0.9    &      7.8     \\  
   55074.30710   &       58.4    &      5.7     \\  
   55102.23907   &       36.0    &      5.4     \\  
   55410.38790   &      464.9    &      4.7     \\  
   55412.38876   &      400.3    &      4.9     \\  
   55413.39281   &      437.3    &      5.1     \\  
   55416.37806   &      400.5    &      5.1     \\  
   55467.22365   &      476.2    &      8.5     \\  
   55468.23303   &      473.3    &      5.8     \\  
   55475.42403   &      489.0    &      4.5     \\  
   55482.20431   &      532.5    &      6.8     \\  
   55491.17869   &      467.0    &      6.4     \\
   55520.28880   &      540.6    &      6.8     \\
   55526.27547   &      495.8    &      7.0     \\
   55530.27249   &      592.9    &      7.0     \\
   55538.04299   &      508.4    &      7.6     \\
   55552.22245   &      549.1    &      6.3     \\
   55566.17674   &      572.6    &      5.9     \\
   55571.16468   &      561.7    &      7.2     \\
   55760.43565   &      483.4    &      5.7     \\
   55789.35501   &      469.6    &      5.2     \\
   55847.39550   &      646.4    &      5.0     \\                    
\enddata                            
\end{deluxetable}                   
                                   
\clearpage                         
\newpage

\begin{deluxetable}{lcc}          
\tablecolumns{3}                                             
\tablewidth{200pt}                                             
\tablecaption{Relative RVs of HD 219415\label{table5}}
\tablehead{\colhead{Epoch [MJD]} & \colhead{RV [m s$^{-1}$]} & \colhead{$\sigma_{RV}$ [m s$^{-1}$]}}
\startdata                                         
   53187.40006    &     -11.7     &       7.6      \\   
   53525.43966     &     -1.5      &      8.8      \\   
   53545.42248     &    -23.1      &      7.6      \\   
   53626.17152     &    -16.5      &      7.7      \\   
   53627.17216     &     -5.7      &      6.2      \\   
   53629.17049     &     -1.1      &      6.6      \\   
   53629.35712     &    -21.3      &      6.7      \\   
   53633.15274     &    -14.5      &      7.3      \\   
   53635.15738     &    -20.4      &      5.7      \\   
   53641.13173     &     -6.3      &      6.3      \\   
   53642.13987     &     -2.4      &      6.9      \\   
   53655.10601     &      6.7      &      6.5      \\   
   53663.25945     &     11.8      &      7.0      \\   
   53892.44493     &      1.0      &      7.4      \\   
   53892.44889     &      5.7      &      6.7      \\   
   53892.45286     &     21.5      &      5.8      \\   
   53895.43792     &     12.3      &      7.2      \\   
   53895.44190     &     18.2      &      6.7      \\   
   53895.44587     &     14.0      &      6.4      \\   
   53899.40469     &     10.5      &      7.1      \\   
   53899.40864     &     13.5      &      7.2      \\   
   53899.41261     &     10.3      &      8.4      \\   
   53901.42493     &      8.6      &     12.8      \\   
   53901.42890     &     22.7      &     10.5      \\   
   53901.43287     &      5.1      &     12.1      \\   
   53904.43579     &     -0.3      &      5.4      \\   
   53904.43976     &     -7.4      &      5.9      \\   
   53904.44373     &      5.1      &      5.5      \\   
   53911.40384     &     10.9      &      5.6      \\   
   53911.40817     &      8.1      &      5.1      \\   
   53911.41250     &      2.3      &      5.0      \\   
   54076.13801     &      2.6      &      6.0      \\   
   54096.08539     &     23.3      &      7.3      \\   
   54347.41372     &     -3.7      &      6.8      \\   
   54375.32882     &      4.5      &      5.1      \\   
   54400.07298     &      3.8      &      5.8      \\   
   54426.18997     &     20.2      &      6.0      \\   
   54454.10736     &     17.8      &      7.7      \\   
   54698.44253     &      7.5      &      6.3      \\   
   54727.35885     &     16.3      &      7.3      \\   
   54755.10432     &     -6.4      &      6.9      \\   
   55049.45800     &     -0.8      &      5.8      \\   
   55084.39427     &    -11.1      &      5.9      \\   
   55110.30134  &   -17.7   &     7.4      \\  
   55199.05491  &    -0.3   &     7.8      \\  
   55444.20009  &   -18.0   &     6.0      \\  
   55468.13783  &   -37.3   &     6.0      \\  
   55730.42347  &    13.8   &     5.7      \\  
   55749.36916  &    -8.1   &     6.1      \\  
   55758.34505  &    -6.1   &     6.2      \\  
   55768.29091  &   -12.5   &     5.6      \\  
   55779.46525  &     1.0   &     6.5      \\  
   55796.43008  &    -5.2   &     4.6      \\  
   55813.22070  &    -4.1   &     5.7      \\  
   55821.17612  &     2.0   &     6.3      \\  
   55830.32469  &     0.5   &     6.2      \\  
   55840.30572  &    -7.1   &     6.0      \\  
\enddata                                                           
\end{deluxetable}

\end{document}